\documentclass[aip,
 amsmath,amssymb,
 reprint,%
]{revtex4-1}

\usepackage{graphicx}
\usepackage{color}
\usepackage{bm}
\usepackage{amsmath}
\usepackage{amssymb}
\usepackage{hyperref}
\usepackage{float}

\def\xcm3{\mbox{cm}^{-3}}

\def\Wxcm2{\mbox{Wcm}^{-2}}

\def\Axm2{\mbox{Am}^{-2}}

\definecolor{red}{rgb}{1,0,0}
\definecolor{blue}{rgb}{0,0,1}

\begin{document}

\preprint{Discontinuity}

\title{Collisionless Rayleigh-Taylor-like instability of the boundary between a hot pair plasma and an electron-proton plasma: the undular mode}

\author{M.~E.~Dieckmann}\email[]{mark.e.dieckmann@liu.se}\affiliation{Department of Science and Technology (ITN), Link\"oping University, 60174 Norrk\"oping, Sweden}
\author{M.~Falk}\affiliation{Department of Science and Technology (ITN), Link\"oping University, 60174 Norrk\"oping, Sweden}
\author{D.~Folini}\affiliation{Ecole Normale Sup\'erieure, Lyon, CRAL, UMR CNRS 5574, Universit\'e de Lyon, 69622 Lyon, France}
\author{R.~Walder}\affiliation{Ecole Normale Sup\'erieure, Lyon, CRAL, UMR CNRS 5574, Universit\'e de Lyon, 69622 Lyon, France}
\author{P.~Steneteg}\affiliation{Department of Science and Technology (ITN), Link\"oping University, 60174 Norrk\"oping, Sweden}
\author{I.~Hotz}\affiliation{Department of Science and Technology (ITN), Link\"oping University, 60174 Norrk\"oping, Sweden}
\author{A.~Ynnerman}\affiliation{Department of Science and Technology (ITN), Link\"oping University, 60174 Norrk\"oping, Sweden}

\date{\today}

\begin{abstract}
We study with a two-dimensional particle-in-cell simulation the stability of a discontinuity or piston, which separates an electron-positron cloud from a cooler electron-proton plasma. Such a piston might be present in the relativistic jets of accreting black holes separating the jet material from the surrounding ambient plasma and when pair clouds form during an X-ray flare and expand into the plasma of the accretion disk corona. We inject a pair plasma at a simulation boundary with a mildly relativistic temperature and mean speed. It flows across a spatially uniform electron-proton plasma, which is permeated by a background magnetic field. The magnetic field is aligned with one simulation direction and oriented orthogonally to the mean velocity vector of the pair cloud. The expanding pair cloud expels the magnetic field and piles it up at its front. It is amplified to a value large enough to trap ambient electrons. The current of the trapped electrons, which are carried with the expanding cloud front, drives an electric field that accelerates protons. A solitary wave grows and changes into a piston after it saturated. Our simulations show that this piston undergoes a collision-less instability similar to a Rayleigh-Taylor instability. The undular mode grows and we observe fingers in the proton density distribution. The effect of the instability is to deform the piston but it cannot destroy it. 
\end{abstract}

\maketitle

\section{Introduction}

Observations\cite{Mirabel92,Siegert2016} of an emission line near 511 keV during a flare of the microquasar V404 Cygni and of pair annihilation radiation in the jets of the microquasar 1E1740.7-2942 evidenced the presence of large clouds of electrons and positrons. This supports the earlier conjecture that microquasars may be an important source of the electron–positron plasma responsible for the bright diffuse emission of annihilation $\gamma$-rays in the bulge region of our Galaxy.~\cite{Siegert2016} Additionally, microquasars could be the origin of the observed megaElectronVolt continuum positron excess in the inner Galaxy.~\cite{Prantzos2011,Dieckmann2019}

Black holes, which accrete material from a companion star and release some of its gravitational energy in the form of jets and radiation, can constitute a microquasar. The accreted material flows onto an accretion disk. Friction heats up the disk. Its inner part can reach temperatures as high as 1~keV. This temperature is inferred from the thermal component of the X-rays. It is emitted by optically thick material, which is most likely that of the disk. It implies that the inner disk is ionized and in a plasma state. The inward flow of plasma and the magnetic field it convects lets the magnetic field accumulate in the inner disk. A magnetized disk is prone to instabilities that can amplify the magnetic field.~\cite{Hawley95,Inchingolo18} 

It has been postulated that the energy release by such instabilities evaporates some of the disk material. A disk corona would form that could account for the observed nonthermal X-ray emissions. The peak energy of the X-rays varies between 100 keV and 200 keV, depending on the disk's state. A temperature this high implies that the plasma is collisionless. Reconnection of magnetic field lines close to the accretion disk may heat the plasma beyond the energy threshold needed to create pairs of electrons and positrons. Pair clouds would thus form close to the reconnection points, which are immersed in the coronal electron-ion plasma (See also the recent review by Yuan et al.\cite{Yuan14}). Thermal motion of particles will let the pair cloud expand. Given that the pair cloud and the ambient coronal plasma are both collisionless and charge-neutral, one may assume that the pair cloud can expand freely. The flow of the pair plasma across the coronal one will, however, trigger plasma instabilities while fluctuations of the electromagnetic fields can scatter particles.~\cite{Dupree63,Bret15} Even in the absence of a background magnetic field, the expanding pair cloud will be coupled to the coronal plasma.~\cite{Dieckmann2018}

How can such a coupling look like? More specifically, can the pair plasma and the coronal plasma mix and form a spatially uniform electron-ion-positron plasma? If this is not the case and both populations remain separated, the plasma of the confined pair cloud may expand along open magnetic field lines of the black hole-accretion disk system (See the related discussion by Dal Pino and Lazarian~\cite{dalpino05}) and leave the vicinity of the accreting black hole. If the separation of both plasmas is maintained even as the pair plasma propagates through the stellar wind of the black hole's companion star or the interstellar medium, it could form jets that are collimated by the inertia of the ions of the ambient plasma. 

\begin{figure}
\includegraphics[width=\columnwidth]{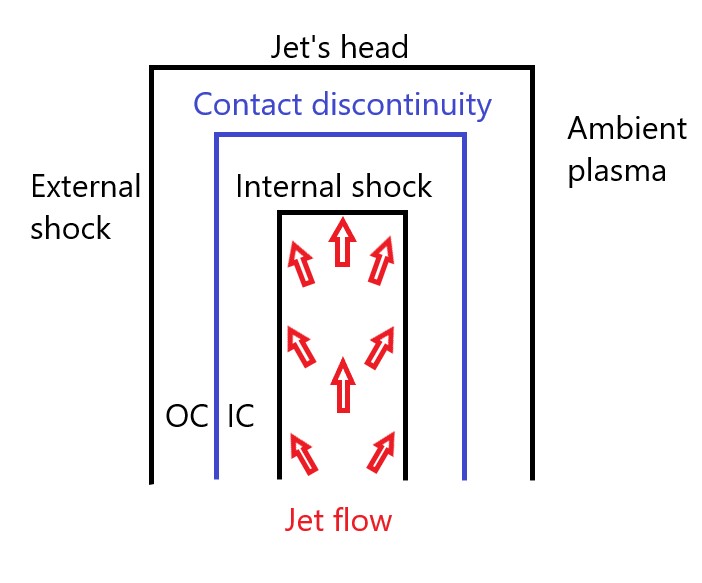}
\caption{Hydrodynamic jet model: A contact discontinuity separates the jet material from the ambient material. It is in touch with the inner cocoon (IC), which contains the jet material that crossed the internal shock. The contact discontinuity is pushed outward by the thermal pressure of the inner cocoon. The moving contact discontinuity lets the surrounding ambient material expand. If this expansion is sufficiently fast then the material of the outer cocoon is separated from the ambient plasma by an external shock. Ambient plasma, which flows across the jet's head, is deflected sideways by the contact discontinuity and remains in the outer cocoon.}
\label{figure1}
\end{figure}

A particle-in-cell (PIC) simulation\cite{Dieckmann2019} demonstrated that a pair cloud expelled the protons of a uniform ambient plasma. A magnetic field was initially aligned with one simulation direction and the mean velocity vector of the pair cloud. A piston in the form of electromagnetic fields grew which expelled protons. This piston acted as the collisionless counterpart of the contact discontinuity in a hydrodynamic jet model,\cite{Marti97,Bromberg11} which is sketched out in Fig.~\ref{figure1}.
The thickness of the piston was comparable to the thermal gyroradius of the cloud particles. Cloud electrons and positrons were confined by this piston and their thermal pressure pushed the piston into the ambient plasma. Electrons of the ambient plasma could not overcome the piston and drifted with it. Their current induced an electric field, which expelled the protons from the interior of the jet. The piston was not planar. The piston's boundary oscillated in space with a wavevector along the expansion direction of the pair cloud suggesting that an instability was at work. 

Pair plasma propagated along the magnetic field of the piston in that simulation. A PIC simulation,\cite{Dieckmann2020} which resolved only the direction perpendicular to the piston and did not give any particle species a net drift along its magnetic field, demonstrated that this drift was not important for stabilizing the piston. However, the process that caused the piston to oscillate in space\cite{Dieckmann2019} could not be determined due to the geometrical constraints. This bending could be caused by two types of instabilities. Pair particles drifting along the piston can trigger a Kelvin-Helmholtz type instability.\cite{Alves2012} Kelvin-Helmholtz instabilities are also known to affect the contact discontinuity between a jet and the ambient material on macroscopic scales.\cite{Perucho2010} An expansion of the piston into the magnetized ambient plasma can trigger a Rayleigh-Taylor instability. In our case the gravitational force~\cite{Winske96,Bret2011,Liu2019} is replaced by the ram pressure, which is excerted by the ions of the ambient plasma on the moving piston. 

Here we present data from a simulation with the same initial conditions as in our previous one\cite{Dieckmann2020} but with a second dimension that is aligned with the background magnetic field. Kelvin-Helmholtz type instabilities cannot develop because the particles are injected in the direction orthogonal to the magnetic field of the piston. We find nevertheless that the piston becomes non-planar, which we thus attribute to a Rayleigh Taylor-type instability. 

Our paper is structured as follows. Section \ref{section2} presents the simulation setup. Results are presented in Section \ref{section3} and discussed in Section \ref{section4}.

\section{Simulation setup}
\label{section2}

Particle-in-cell (PIC) simulation codes solve Amp\`ere's and Faraday's laws on a numerical grid. A plasma species $i$ composed of particles with the charge $q_i$ and mass $m_i$ is approximated by an ensemble of computational particles (CPs). Each CPs must have the same charge-to-mass ratio $q_i/m_i$ as the represented species. The electric field $\mathbf{E}$ and the magnetic field $\mathbf{B}$ are defined on the grid. Their values are interpolated to the position $\mathbf{x}_j$ of the $j^{th}$ CP and the particle momentum $\mathbf{p}_j = m_j\Gamma_j \mathbf{v}_j$ ($c, m_j, \Gamma_j$ : light speed, mass and relativistic factor $\Gamma_j = {(1-\mathbf{v}_j^2/c^2)}^{-1/2}$ of CP $j$) is updated with a discretized form of the Lorentz force equation. The current $\propto \mathbf{v}_j$ of the CP is interpolated onto the grid. Summing up the current contributions of all CPs yields the macroscopic plasma current $\mathbf{J}$, which updates $\mathbf{E}$ and $\mathbf{B}$ via Amp\`ere's law. We use the EPOCH code.\cite{Arber2015}

Initially, the simulation box is filled with a spatially uniform ambient plasma, which consists of electrons and protons with the correct particle mass ratio $m_p/m_e = 1836$. Each plasma species has the density $n_0$ and the temperature $T_0 =$ 2 keV. The ambient plasma corresponds to the coronal plasma at the jet source region and to the stellar wind of the companion star at larger distances from this region. Our value for $T_0$ allows us to use a coarse grid without triggering self-heating instabilities\cite{Arber2015} while the ambient plasma is still cold compared to the temperatures that are reached at later times.

Densities are normalized to $n_0$. Time and space are normalized to the inverse of the proton plasma frequency $\omega_{pi}={(n_0e^2/\epsilon_0 m_p)}^{1/2}$ ($e, \epsilon_0, \mu_0$: elementary charge, vacuum permittivity and permeability) and to the proton skin depth $\lambda_{pi}=c/\omega_{pi}$. Unless stated otherwise, electric and magnetic fields are normalized to $m_p c \omega_{pi} / e$ and $m_p \omega_{pi} / e$. A magnetic field with the amplitude $B_0 =$ 0.0021 is aligned with $y$. The simulation box length $L_x = 26.4$ along $x$ is resolved by 9000 grid cells while 2250 grid cells resolve its length $L_y = 6.6$ along $y$. Boundary conditions are periodic in all directions. We evolve the simulation during $0 \le t \le t_{max}$ with $t_{max}=$ 190 and use for this purpose $1.05 \times 10^5$ time steps $\Delta_t$. Each species of the ambient plasma is resolved by $8.1 \times 10^8$ CPs.  

We want to model a piston that corresponds to the contact discontinuity in a hydrodynamic jet model. We consider in Fig.~\ref{figure1} a horizontal slice of the jet in which the discontinuity is aligned with the vertical direction. This piston is in contact with the ambient plasma on one side. We find on its other side pair plasma that has crossed the internal shock and entered the jet's inner cocoon. This shocked pair plasma has a high temperature and should be close to a thermal equilibrium. Its mean speed equals the nonrelativistic lateral expansion speed of the contact discontinuity. The piston should grow self-consistently from an interaction between the ambient plasma and a pair plasma with initial conditions that are easy to implement and lend themselves to parametric studies. 

\begin{figure}
\includegraphics[width=\columnwidth]{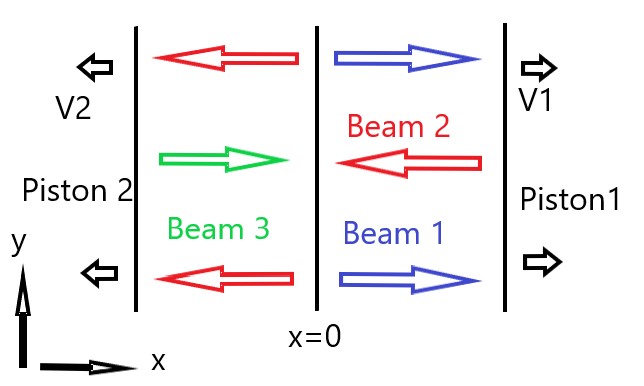}
\caption{Simulation setup: Electron-positron pairs are injected at the boundary $x=0$ in the direction of increasing $x$ forming beam~1 (blue). This pair cloud is reflected by piston~1 and transfers momentum and energy to it. Piston~1 expands with the speed $v_1$ to increasing $x$. The pair particles are reflected. The returning beam~2 (red) does not interact strongly with beam~1 due to its high temperature. It crosses the boundary at $x=0$. It propagates until it reaches piston~2, where it is reflected again. The momentum transfer lets piston~2 expand at the speed $v_2$ towards negative $x$. Beam~3 (green) crosses the boundary $x=0$ and is mixed with beam~1. The background magnetic field is aligned with $y$.}
\label{figure2} 
\end{figure}

For this purpose, electrons and positrons are injected at $x=0$ with the mean speed $v_0/c=$ 0.75 along increasing $x$ forming beam 1 in Fig.~\ref{figure2}. 
Each species has the number density $n_0$ measured in the simulation box frame. At every time step, 270 000 CPs are injected and distributed evenly over both cloud species. Their velocity distribution in the rest frame of the injected cloud is a non-relativistic Maxwellian with the temperature $T_c =$ 100 keV. The thermal speed of the pair cloud $v_c = {(k_B T_c / m_e)}^{1/2}$ ($k_B$ : Boltzmann constant) is $v_c / v_0 \approx 0.6$. This high thermal speed reduces the impact of beam instabilities between the injected pair plasma and the one that has been reflected by $B_0$ (beam 2 in Fig.~\ref{figure2}). The injected pair cloud will thus impose a ram pressure on the ambient plasma, which is almost constant in space and time. The reflected returning pair plasma will cross the boundary and pile up another piston on the opposite side. Multiple reflections of the pair clouds by both pistons (e.g. beam 3 in Fig.~\ref{figure2}) will increase the density of the pair plasma in time until an equilibrium is reached between its pressure and the ram pressure of the expanding ambient plasma. 

In Fig.~\ref{figure1}, the pair plasma on either side of the simulation boundary $x=0$ would be located in the inner cocoon close to the discontinuity. In other words, we cut out the jet flow and the internal shocks and stick both segments together. Our simulation will show that permanently injecting pair plasma at the boundary ensures the expansion of both pistons at a uniform speed.

\section{Simulation results}
\label{section3}

The first subsection examines the growth and saturation of the electromagnetic fields that constitute the piston. The second subsection investigates how the characteristic wavelength of the piston's spatial oscillations couple from electron skin depth-scales to proton skin depth-scales. The final distribution at $t_{sim}=190$ is addressed by the third subsection.

\subsection{Early time}

\begin{figure*}
\includegraphics[width=\textwidth]{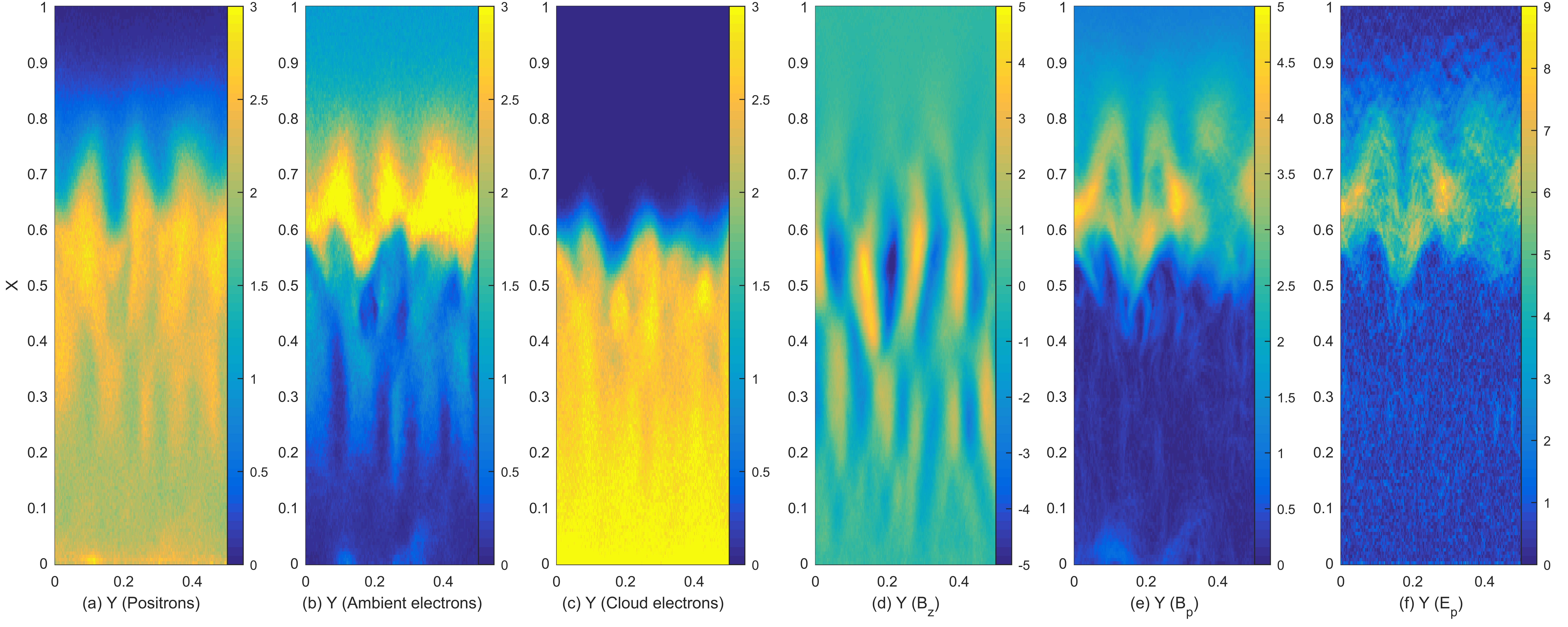}
\caption{The plasma at $t=5$: The density distributions of the positrons, ambient electrons and cloud electrons are shown in panels (a-c), respectively. The out-of-plane magnetic field $B_z$ and the in-plane component $B_p = {(B_x^2+B_y^2)}^{1/2}$ are displayed in panels (d) and (e). Both are normalized to $B_0$. Panel (f) shows the normalized in-plane electric field $E_p= {(E_x^2+E_y^2)}^{1/2}$.}
\label{figure3}
\end{figure*}

Figure~\ref{figure3} shows the plasma and field distributions at the time $t=5$. Pair particles, which were injected with $v_0/c$ = 0.75 at $t=0$, have propagated  the distance 3.75.  
Positrons have their density peak at $x\approx 0.55$ in Fig.~\ref{figure3}(a). Their density decreases rapidly for larger $x$ and goes to zero at $x\approx 1 \ll 3.75$; by this time, leptons completed 2.7 gyroperiods in the initial magnetic field $B_0$ and their expansion is thus not free. The front of the positron cloud is rippled despite it being injected with a uniform density. On average, the positron density has a value just above the sum of the densities of the injected and reflected pair beams (beams 1 and 2 in Fig.~\ref{figure2}); the pair plasma has not yet been compressed. Figures~\ref{figure3}(b,~c) depict the densities of the ambient electrons and the cloud electrons. Ambient electrons neutralize the charge density of the positrons at the cloud front. Hardly any ambient electrons are found in the interval $x<0.5$. The cumulative electron density exceeds that of the positrons, which is seen best in the interval $x<0.3$. This excess negative charge density balances that of the protons.

Oscillations of $B_z(x,y)$ in Fig.~\ref{figure3}(d) evidence a Weibel-type instability. Weibel-type means in this context that interactions of electrons and positrons via their microscopic currents collimate the particles into current channels in the $x,y$ plane. These channels are separated by magnetic fields that point along $z$. In what follows, this generic term incorporates both the Weibel instability in its original form,~\cite{Weibel1959} its extensions to (un-)magnetized pair plasma~\cite{Schlickeiser2010} and the filamentation instability of counter-streaming lepton beams.~\cite{Bret2010} For values $0.2 \le x \le 0.4$, the magnetic field is sustained by the separation of electrons and positrons of the injected cloud. This can be seen in particular in the interval $y \le 0.2$ in Figs.~\ref{figure3}(a, c), where the density peaks of the positrons and cloud electrons are interlaced. The density distribution of the ambient electrons that remain behind the cloud front follows closely that of the positrons like for example for $0.2 \le x \le 0.5$. 

Figure~\ref{figure3}(e) demonstrates that the background magnetic field, which is oriented along $y$, has been evacuated from the interval $x<0.5$ by the expanding cloud. The likely mechanism is the cloud particle's diamagnetic current. Electrons and positrons rotate in opposite directions around $B_0$ and their current does not cancel out at the cloud front. This net current depletes the background magnetic field within the cloud and piles it up ahead of it. Figure~\ref{figure3}(e) confirms that the field has been piled up in the interval ahead of the cloud electrons. It is correlated with a strong in-plane electric field in Fig.~\ref{figure3}(f). Ambient electrons, which drift with $B_p$ along $x$, induce this field. The in-plane electric field $E_p$, which is polarized on average along $x$, is strong enough to accelerate protons. Protons gain speed and are compressed in this direction; a solitary wave grows as we show below. 

\begin{figure}
\includegraphics[width=\columnwidth]{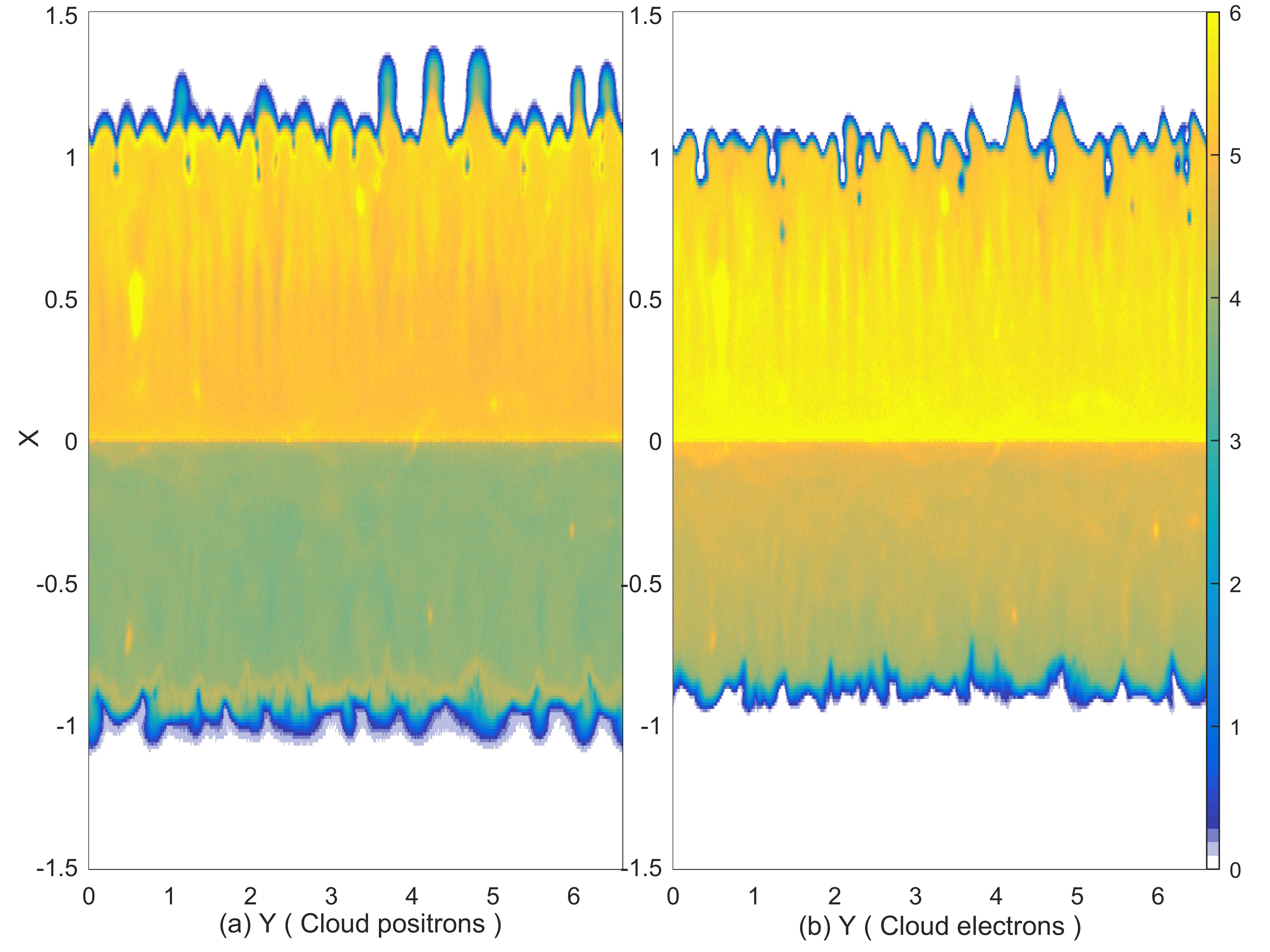}
\caption{Pair cloud density at $t=20$: Panel (a) shows the positron density and panel (b) that of the electrons. Both panels use the same color scale. Multimedia view:}
\label{figure4}
\end{figure}

Figure~\ref{figure4} depicts the densities of both species of the pair cloud at the time $t=20$.
Their density is larger in the half space $x>0$ than in $x<0$. Cloud particles are reflected by an expanding non-planar front. They are scattered into a wide angular range, which heats them up along $y$, and they lose momentum to the moving boundary. Cloud particles are thus slowed down along $x$ and compressed in the half-space $x>0$, which reduces the number of particles that cross the boundary towards negative $x$. We relate this observation to the jet model in Fig.~\ref{figure1}: Once both pistons are located sufficiently far from the boundary, we could split the simulation box at the boundary $x=0$ and consider the pair plasma on each side as the one we find in the inner cocoon of the jet. The pair plasma on each side of the boundary $x=0$ belongs in this case to an inner cocoon with a different temperature and density. With respect to the model in Fig.~\ref{figure1}, this could correspond to jets with different strengths of the internal shocks or to different flow speeds of the pair plasma in the jet flow channel prior to the shock crossing; faster flows lead to a larger temperature and density of the pair plasma downstream of the internal shock. We can thus study with one simulation the interaction of an ambient plasma with two shocked pair plasmas with different thermal pressures and test how robust the processes are that lead to the formation and evolution of the piston. Both cloud species reveal sharp boundaries at large values of $|x|$ with the positron boundary being located at larger $|x|$. The $x-$position of the boundaries varies with $y$. Fingers have grown in the positron density distribution in Fig. \ref{figure4}(a) at $3.5 \le y \le 5.5$ and $x\approx 1.2$. 

\begin{figure}
\includegraphics[width=\columnwidth]{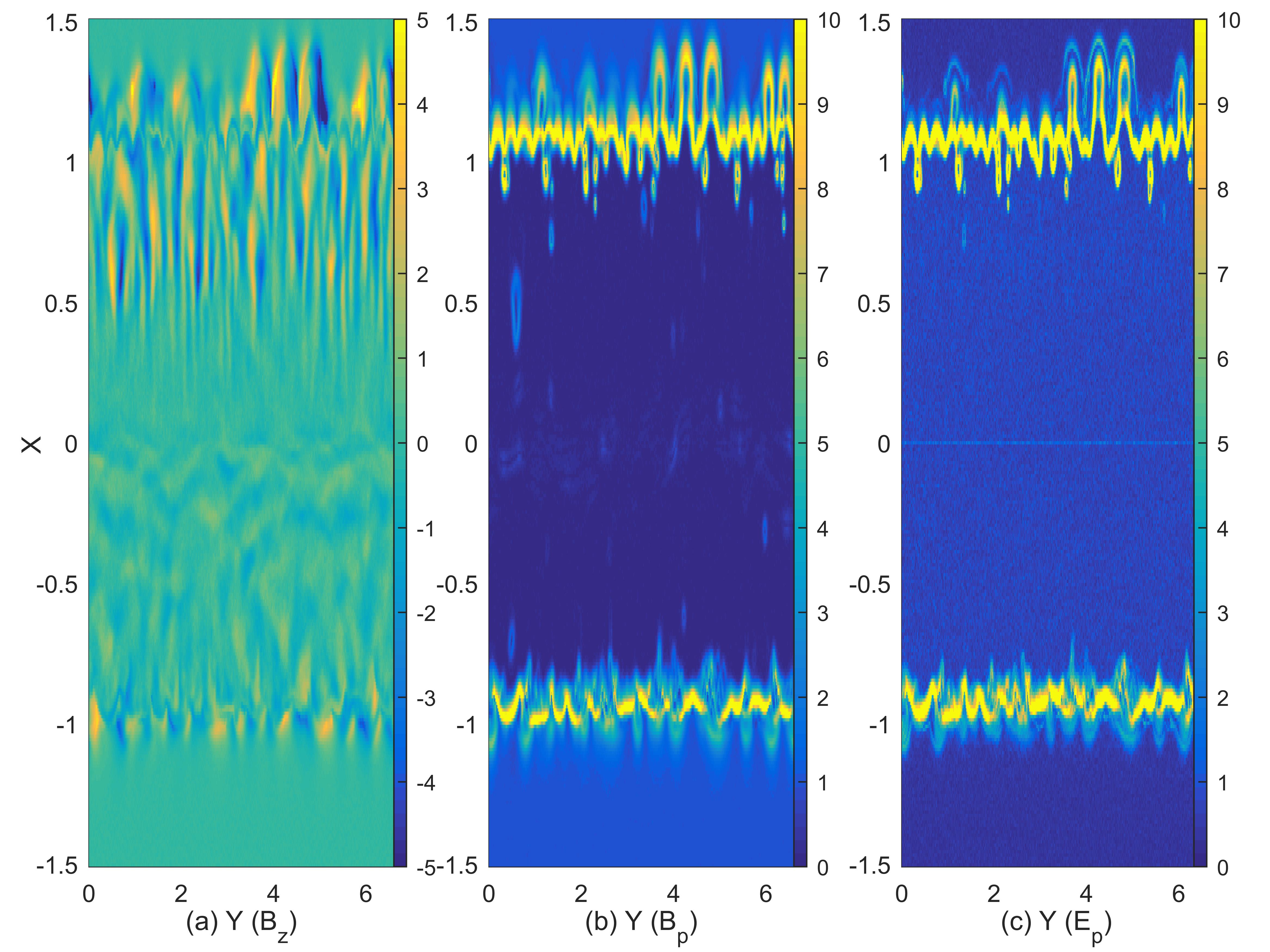}
\caption{Electromagnetic fields at $t=20$: Panel (a) shows the out-of-plane magnetic field $B_z$. The in-plane magnetic field $B_p = {(B_x^2+B_y^2)}^{1/2}$ is shown in panel (b). Both are expressed in units of $B_0$. Panel (c) shows the normalized in-plane electric field $E_p={(E_x^2+E_y^2)}^{1/2}$. Multimedia view:}
\label{figure5}
\end{figure}

Figure~\ref{figure5} shows the out-of-plane magnetic field $B_z$, the in-plane magnetic field $B_p$ and the in-plane electric field $E_p$ at $t=20$.
According to Fig.~\ref{figure5}(a), the out-of-plane magnetic field is strongest ahead of the boundary in the half space $x>0$, which is located at $x\approx 1.1$. Oscillation amplitudes go through a minimum at the boundary and increase again for $x<0.9$. Oscillations of $B_z$ with a lower amplitude are also observed at $x\approx -1$ just ahead of the boundary in the half-space $x<0$. The fingers in the positron density distribution at $x\approx 1.2$ and $y\approx 4.5$ in Fig.~\ref{figure4}(a) are enclosed by strong electromagnetic fields. Figure~\ref{figure5}(b) confirms that the expanding pair cloud continues to expel the background magnetic field.  

\begin{figure}
\includegraphics[width=\columnwidth]{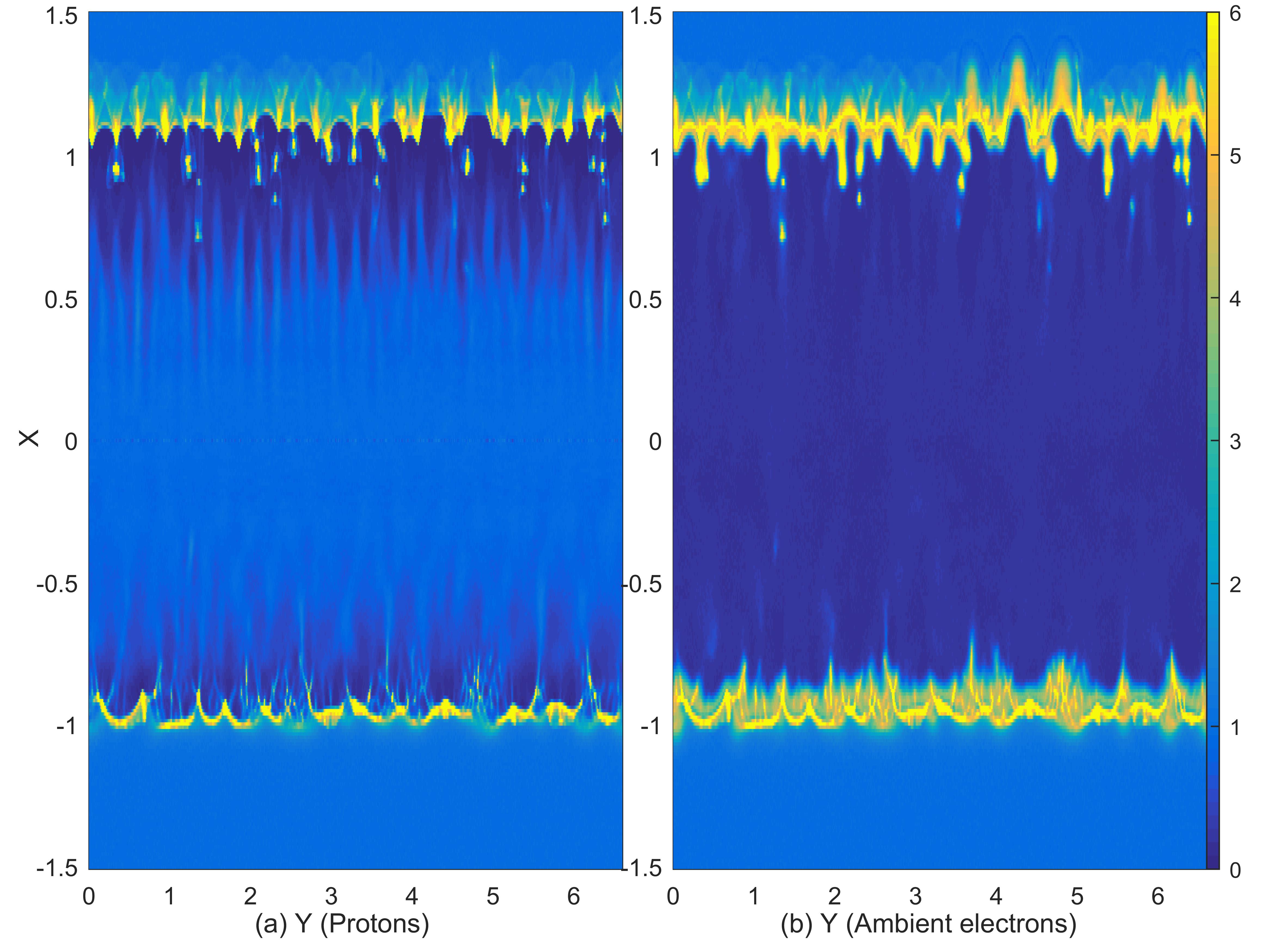}
\caption{Density distribution of the ambient plasma at the time $t=20$: Panel (a) shows the proton density and panel (b) that of the ambient electrons. Both panels use the same color scale. Multimedia view:}
\label{figure6}
\end{figure}

Protons with a density close to $n_0$ are found in Fig.~\ref{figure6}(a) for $|x| \le 0.7$. 
Their presence lets the density of the cloud electrons exceed that of the positrons for $|x| \le 0.9$ in Fig.~\ref{figure4}. The proton density does not change across $x=0$ in Fig.~\ref{figure6}(a) and they can thus not cause a change in the plasma density at this location. Their high mobility lets leptons diffuse from the region $x>0$ in Fig.~\ref{figure4} with a high density to the one $x<0$ with a low one. The electron and positron density jumps across $x=0$ in Fig.~\ref{figure4} are comparable in size and a similar number of particles of each species diffuse across the boundary. No large net charge can build up and the ambipolar electric field at $x=0$ in Fig.~\ref{figure5}(c) remains weak. 

High-density bands in the ambient plasma are located in Fig.~\ref{figure6} at $x\approx -1$ and $x\approx 1.1$ with a density distribution that is not uniform along $y$. Protons clump together in particular at the boundary $x\approx 1.1$. Hardly any protons are left in the interval $0.9 \le x \le 1.1$, which evidences that the in-plane electromagnetic field at $x\approx 1.1$ in Fig. \ref{figure5} has become strong enough to sweep them out; the piston has formed. Most ambient electrons have also been expelled from the interval $-0.8 \le x \le 0.9$ by the piston. Their distribution at $x\approx 1.2$ and $y \approx 4.5$ outlines the fingers in the positron distribution in Fig.~\ref{figure4}(a). 

Both electron species are separated by the piston at $x\approx 1.1$. Oscillations of $B_z$ at $x>1.1$ in Fig.~\ref{figure5}(a) must thus be caused by a Weibel-type instability between the positrons and the ambient electrons. Those in the interval $0.5 \le x \le 1$ can only be tied to a Weibel-type instability between the cloud particles. The latter develops because the velocity spread of the injected and returning cloud particles is larger along $x$ than along $y$ due to the mean speed we give them at the injection line $x=0$. The wavevector of Weibel-type modes is aligned with the cool direction, which is here $y$ and matches the observed modulation of $B_z$. The instabilities ahead and behind the pistons have not been observed previously in this form.\cite{Dieckmann2019} We can attribute that to the motion of pair particles along $\mathbf{B}_0$ in the earlier simulation. Reflected leptons with a velocity component along $\mathbf{B}_0$ do not flow antiparallel to the inflowing ones. Here, the Weibel-type instability between the counter-streaming leptons perturbs the piston, which provides the seed for secondary instabilities.

Insight into how the piston is sustained is provided by the phase space density distributions of the individual plasma species. These are $f_e(x,y,E)$, $f_p(x,y,E)$ and $f_i(x,y,E)$ for electrons (ambient and cloud electrons), positrons and protons, respectively. Kinetic energies $E$ are expressed in units of MeV. In what follows, we display the square root of these densities in order to resolve adequately their high energy tails. We normalize electron and positron distributions to the initial peak density of the ambient electrons and the proton distribution to its initial peak value.  

\begin{figure*}
\includegraphics[width=\textwidth]{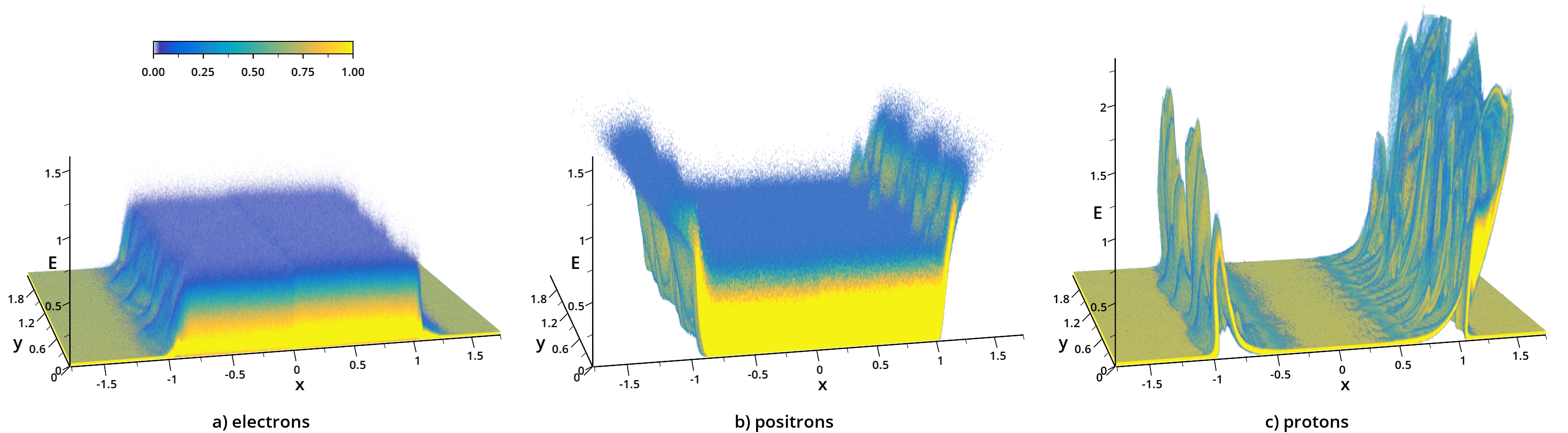}
\caption{Energy distributions at the time $t=20$. Energy $E$ is expressed in units of MeV. Panel (a) shows $f_{e}(x,y,E)^{1/2}$ of the electrons, which was summed over ambient and cloud electrons. Panel~(b) shows the energy distribution of the positrons $f_{p}(x,y,E)^{1/2}$ and (c) renders $f_i(x,y,E)^{1/2}$ of the protons. Multimedia view:}
\label{figure7}
\end{figure*}

Figure~\ref{figure7} shows these distributions at the time $t=20$, which were rendered with Inviwo.\cite{Inviwo}
Both lepton distributions are uniformly distributed in space within $0 \le x \le 0.9$ and within $-0.9 \le x \le 0$. The phase space density distribution of the electrons falls off at large $|x|$ and goes over into the distribution of the background electrons. Positrons at the front of the clouds have a larger energy. Furthermore, the minimum energy of the positrons at the front of the injected pair cloud increases with $|x|$  while the maximum energy of the electrons decreases with increasing $|x|$. This difference is caused by the piston's in-plane electric field. Positrons move ahead of the electric field band where they rotate in the background magnetic field, which keeps their energy unchanged. Hence, more energetic positrons make it farther upstream than low-energy positrons explaining the forward-tilt of the front. They rotate in the magnetic field, which drives a current in the z-direction that is not cancelled out by an electronic one. The resulting net current along $z$ amplifies $B_p$. The in-plane electric field also accelerates protons. Protons at the boundary at $x \approx 1$ reach a higher peak energy than those at $x\approx -0.9$ and the boundary in the interval $x>0$ has advanced farther. The faster propagation speed of the piston in the half-space $x>0$ and its more powerful proton acceleration is sustained by the larger thermal pressure of the lepton cloud in this half-space. 

Figure~\ref{figure7} (multimedia view) reveals how the piston forms. The in-plane electric field induced by the trapped ambient electrons accelerates protons. Protons in the half-space $x>0$ ($x<0$) obtain positive (negative) speeds. They are piled up and a solitary density wave grows. It develops over a time scale much shorter than a proton gyro-period. However, since the proton density pulse is accompanied by a localized peak of the magnetic pressure and because it propagates across the magnetic field, we may interpret it as a solitary fast magnetosonic wave. A driving electric field $E_p \neq 0$ implies that this wave is not a soliton. Electric fields close to a soliton are self-generated by changes in the thermal and magnetic pressures. Unlike solitons, driven solitary waves can change their amplitude and this is what we observe. The solitary wave saturates once the ambipolar electric field, which is caused by the large variations of the thermal and magnetic pressures, becomes large enough to reflect protons back upstream; the solitary wave breaks. Once the solitary wave has saturated it becomes the piston that keeps apart positrons and protons. 

Figure~\ref{figure7}(multimedia view) demonstrates that only electrons lose a substantial fraction of their energy once protons pick up speed. This different response of electrons and positrons to proton acceleration is visible at $t=20$. Positrons reach higher energies than electrons across the interval $-0.8 \le x \le 0.8$ where each species is close to a thermal equilibrium. There is a jump along the energy axis at $x=0$. It arises because the injected cloud particles experience a loss of energy when they are reflected by the moving piston in the interval $x>0$.\cite{Dieckmann2020}

\subsection{Coupling across length scales}

So far, spatial oscillations of the piston were seeded by Weibel-type instabilities. Their wavelength was limited to a few electron skin depths (one electron skin depth equals $\sqrt{m_e/m_p}$ in our normalization) as can be seen from Fig.~\ref{figure3}. In our simulation, light cloud particles are pushing against protons, which can yield a Rayleigh-Taylor instability. Since the piston's magnetic field is oriented in the simulation plane and orthogonal to the cloud's expansion direction, this instability involves the undular mode. Undular modes have a wavevector that is parallel to the magnetic field of the piston. Figure~\ref{figure4} revealed fingers in the positron density distribution that were reminiscent of a Rayleigh-Taylor instability. 

\begin{figure*}
\includegraphics[width=\textwidth]{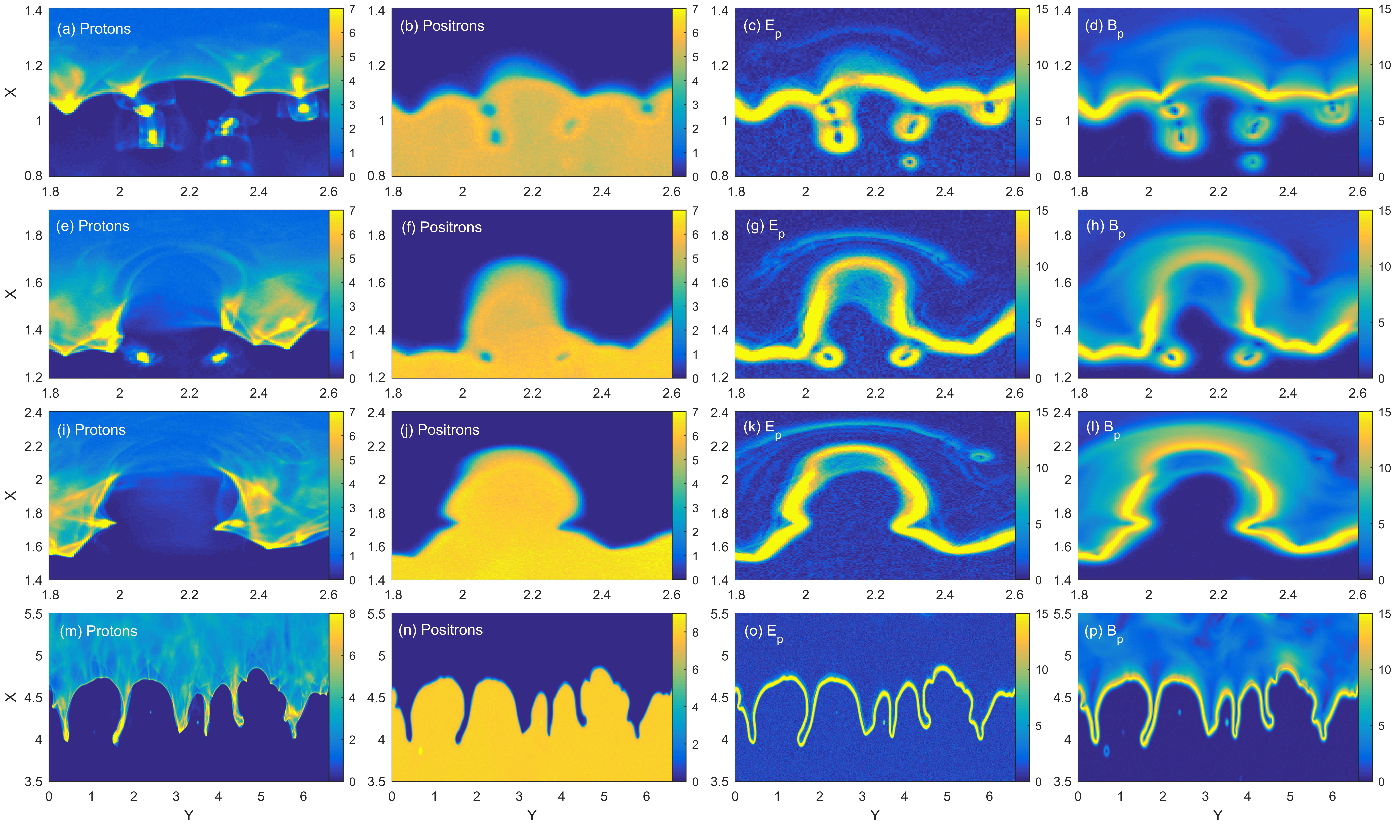}
\caption{Densities of positive charge carriers and in-plane electromagnetic fields at selected times. Panels (a-d) show the density distributions of protons, positrons and the amplitudes of the in-plane electric field $E_p = (E_x^2+E_y^2)^{1/2}$ and magnetic field $B_p = (B_x^2+B_y^2)^{1/2}$ at the time $t$ = 20, respectively. Panels (e-h) display the density distributions of protons and positrons, $E_p$ and $B_p$ at the time $t=30$. Panels (i-l) depict the density distributions of protons and positrons, $E_p$ and $B_p$ at the time $t=40$. The density distributions of protons, positrons, $E_p$ and $B_p$ at $t=130$ are shown in panels (m-p). Multimedia view:}
\label{figure8}
\end{figure*}

Positrons and protons are unique markers for the light and heavy fluids, respectively. A Rayleigh-Taylor instability involves $B_p$ and $E_p$. Figure~\ref{figure8} tracks the aforementioned quantities close to a growing finger.
At the time $t=20$, the piston has fully developed and separates protons and positrons. Oscillations of the piston have a wavelength between 0.1 and 0.2. Figure~\ref{figure8}(d) reveals that $B_p$ is depleted at $y\approx 2.15$ and $x=1.15$. The magnetic field lines of the piston have spread out over $x$. Positrons in Fig.~\ref{figure8}(b) bulge out into the region where the magnetic field has been weakened. Arcs in the proton distribution at $y\approx$ 2.1 and 2.25 and $x\approx 1.2$ mark parts of the front of the expanding positrons. These protons were pushed out by $E_p$ in Fig.~\ref{figure8}(c)(Multimedia view). A depletion of $B_p$ and an expansion of positrons takes place also at $y\approx 1.95$ and $y\approx 2.4$. These three values of $y$ correspond to locations where the piston is concave with respect to the positrons. In other words, a deformation of the piston into the heavy fluid weakens it. Positrons overcome the weakened piston and flow upstream. Protons have a relatively low density at these locations. They accumulate where the piston is convex relative to the positrons. Those parts of the piston where protons accumulate offer more resistance to the expansion of the pair cloud and this increased inertia amplifies the boundary oscillation triggering a thin shell instability.~\cite{Dieckmann2015} The piston is strong enough to confine most of the protons. Miniscule proton clouds crossed the piston at positions $y$ where protons accumulated, for example at $x\approx 1.02$ and $y\approx 2.05$.

Figures~\ref{figure8}(e-h) have been sampled at the time $t=30$. A positron finger in the interval $2 \le y \le 2.3$ extends from $x$ =1.3 to 1.7. Its boundary at large $x$ is again correlated with dilute arcs in the proton density distribution. These arcs are now closed. However, the majority of the protons has not yet been expelled from the interval occupied by the positrons. The piston is still sustained by the positron current at the front of the positron finger. It is immersed in a weaker diffuse magnetic field patch in Fig.~\ref{figure8}(h), which spreads out over a large $x$-interval. Electric fields bands ahead of the piston at $x \approx 1.8$ outline the front of the diffuse magnetic patch, which is still strong enough to trap ambient electrons that induce in turn the electric field. Apparently, electromagnetic fields also develop in the interval close to $y\approx 2.15$ and $x\approx 1.6$ behind the piston where positrons and protons still coexist. Figure ~\ref{figure8}(Multimedia view) demonstrates that the effect of this field is to accelerate protons to increasing $x$ and to speeds in excess of that of the piston. 

Figures~\ref{figure8}(i-l) are sampled at $t=40$. The positron finger has kept its length and width almost unchanged. Its expansion stretches the magnetic field, which creates a magnetic tension force that counteracts a further expansion of the finger. The electromagnetic field behind the piston has reduced the proton density in the interval occupied by the positrons. Proton clusters are located in Fig.~\ref{figure8}(i) at $x$=1.75 with values $y$ =1.97 and 2.3. Their positive charge expels the positrons in these intervals in Fig.~\ref{figure8}(j), which gives the positron finger a mushroom shape. This shape is typical for the non-linear stage of the Rayleigh-Taylor instability. However, it is the density distribution of the heavy species that forms the mushroom in the hydrodynamic Rayleigh-Taylor instability and not the light one. The piston can also not separate completely protons and positrons in our simulation. Some differences thus exist between the Rayleigh-Taylor instability and the one that deforms the piston in our simulation for $20 \le t \le 40$. 

Figures~\ref{figure8}(m-p) demonstrate that the piston has pleated at the time $t=130$. Its oscillations span an interval with the width 0.6 along $x$ and their wavelength can be as large as one proton skin depth. The piston has, however, remained stable and the protons that were trailing it previously were pushed ahead of it. We observe a complete separation between broad positron fingers and narrow proton fingers. A propagation of the piston to increasing $x$ and the polarisation of the electric field at the piston implies that the electric field funnels protons into the fingers. They propagate to the end of the finger where they are reflected. Their large number density maximizes the momentum transfer from the protons to the piston at this end point, which may elongate further the proton finger. An almost complete separation of the heavy and light fluids in Figs.~\ref{figure8}(m, n) and the growth of proton fingers is what we expect from a hydrodynamic Rayleigh-Taylor instability. However, it is not an exact counterpart of the hydro-dynamic Rayleigh-Taylor instability given that this instability was triggered by the bulging of positrons across a weakened piston rather than by a gradual growth of the oscillation of a piston that keeps positrons and protons separated at all times. We thus refer to it as Rayleigh Taylor-like instability.

\subsection{Distribution at the simulation's end}

\begin{figure*}
\includegraphics[width=\textwidth]{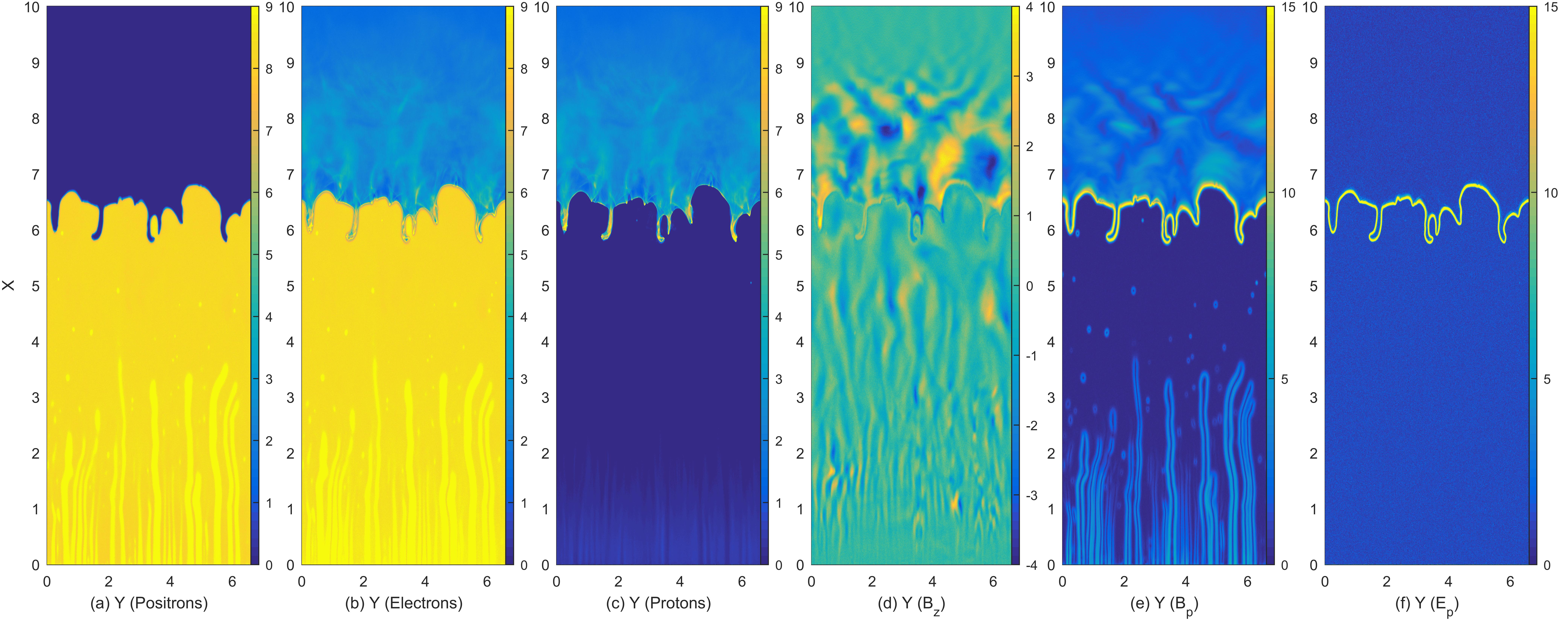}
\caption{Distributions of the particle density and electromagnetic fields at the time $t_{sim}=190$ in the half-space $x>0$: Panels (a-c) show the density distribution of the positrons, that of the electrons summed over ambient and cloud electrons and that of the protons. Panels (d, e) show the distributions of the magnetic $B_z$ component (out-of-plane) and of $B_p$ (in-plane) expressed in units of $B_0$. Panel (f) shows the distribution of the in-plane electric field $E_p$. Multimedia view:}
\label{figure9}
\end{figure*}

We examine here how the piston evolves after $t=130$. More specifically, we want to determine if the magnetic tension force can prevent a continuing elongation of the proton fingers. Figure~\ref{figure9} presents the relevant plasma densities and field distributions at $t_{sim}=190$. 
Protons and positrons continue to be separated, which demonstrates that the piston is stable. The front of the pair cloud maintained its shape. Electrons and positrons have a uniform density behind the piston. Elongated density striations are visible at lower $x$. Figure~\ref{figure9}(Multimedia view) reveals that they start growing at the boundary and expand from there to larger $x$. These striations are confined by an in-plane magnetic field in Fig.~\ref{figure9}(e). 

We observed similar structures~\cite{Dieckmann2019} when we injected electrons and positrons at a reflecting boundary. They appear after several tens of inverse proton plasma frequencies and their source is thus the proton distribution close to the boundary $x=0$. A comparison of Fig.~\ref{figure6}(a) and Fig.~\ref{figure9}(c) reveals that proton density filaments have grown between $t$ = 20 and 190 with a diameter along $y$ that is comparable to that of the striations. Positively charged proton filaments close to the boundary and the need to maintain quasi-neutrality enforces a rearrangement of the injected electrons and positrons. This rearrangement yields the growth of a net current and a magnetic field. The latter remains weak and the striations are separated from the piston by 3 proton skin depths. Their effects on the piston should thus be negligibly small.

The proton fingers in Fig.~\ref{figure9}(c) retained their shape and length compared to those in Fig.~\ref{figure8}(m) while the piston as a whole has propagated for a distance $\approx$ 2 along $x$. The undular mode of the Rayleigh Taylor-type instability was thus either stabilized by the magnetic tension force or it evolves on time scales much longer than a few tens of inverse proton plasma frequencies. 

Structures in the out-of-plane magnetic field in the interval $x<6$ in Fig.~\ref{figure9}(d) are driven by a nonthermal distribution of the cloud behind the piston. More powerful magnetic field oscillations ahead of the piston in Fig.~\ref{figure9}(d) may be driven partially by a Weibel-type instability. Another source is the current due to protons that were scattered into a wide angular range by their reflection by the nonplanar piston. With the exception of the striations, the interval $0 \le x \le 6$ is free of any in-plane magnetic field $B_p$ and electric field $E_p$.

\begin{figure*}
\includegraphics[width=\textwidth]{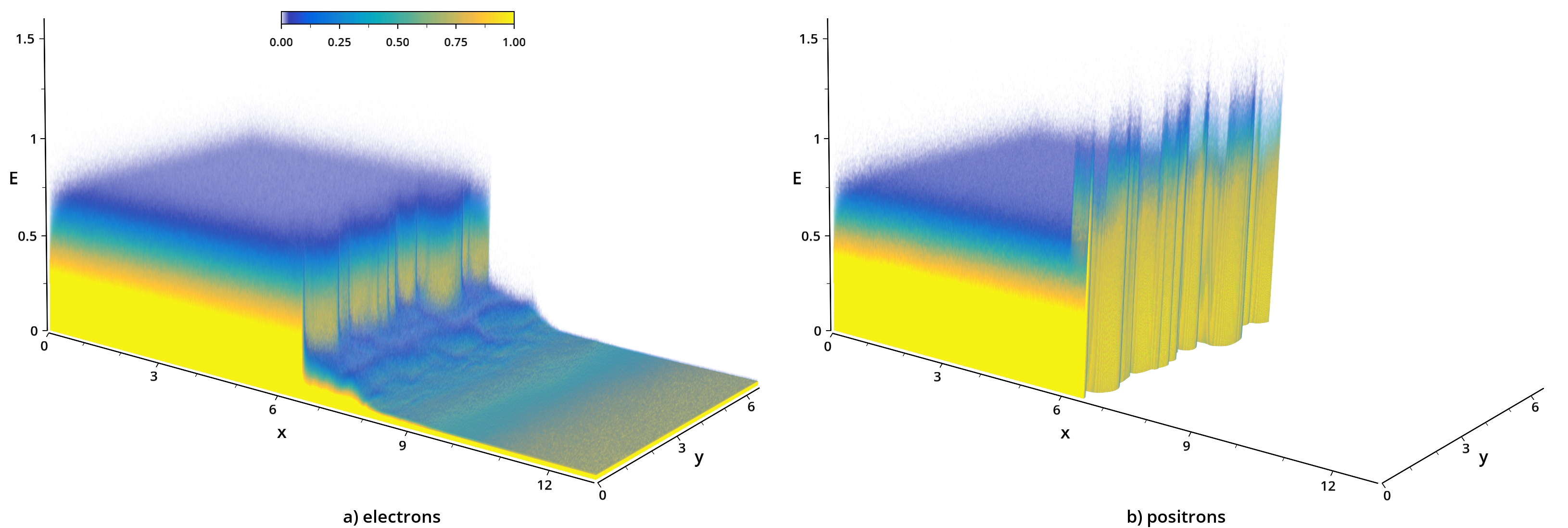}
\caption{Energy distributions of the leptons at the time $t_{sim}=190$: Panel (a) shows the distribution $f_e(x,y,E)^{1/2}$ of the electrons, which includes ambient and cloud electrons. Panel (b) shows the positron distribution $f_p(x,y,E)^{1/2}$. Energies $E$ are expressed in units of MeV. Multimedia view:}
\label{figure10}
\end{figure*}

What is the energy distribution of the plasma particles at the time $t_{sim}$? Figure~\ref{figure10} shows $f_e(x,y,E)^{1/2}$ of all electrons in the half-space $x>0$ and its positronic counterpart $f_p(x,y,E)^{1/2}$ in the same normalization as in Fig.~\ref{figure7}.
Featureless distributions of both cloud species for $x<6$ demonstrate that they are close to a thermal equilibrium. Typical electron energies are below those of positrons as we had already observed in Fig.~\ref{figure7}; protons gain energy at the expense of that of the electrons. The striations in Fig.~\ref{figure9}(a, b) have no effect on the energy distribution of the particles. Positrons are again accelerated close to the piston while electrons are decelerated. Electron heating takes place in the interval $6.5 \le x \le 8$ ahead of the piston. We attribute this heating to the increased proton density in this interval (See Fig.~\ref{figure9}(d)). Thermal diffusion of electrons lets the interval with a larger proton density go on a positive potential relative to the far upstream region. Upstream electrons that approach this interval are accelerated by the ambipolar electric field, which maintains the potential jump, towards the piston and they gain energy. The slow spatial change of the proton density yields an amplitude of the ambipolar electric field that is not large enough to be detectable in Fig.~\ref{figure9}(f).  

\begin{figure*}
\includegraphics[width=0.65\textwidth]{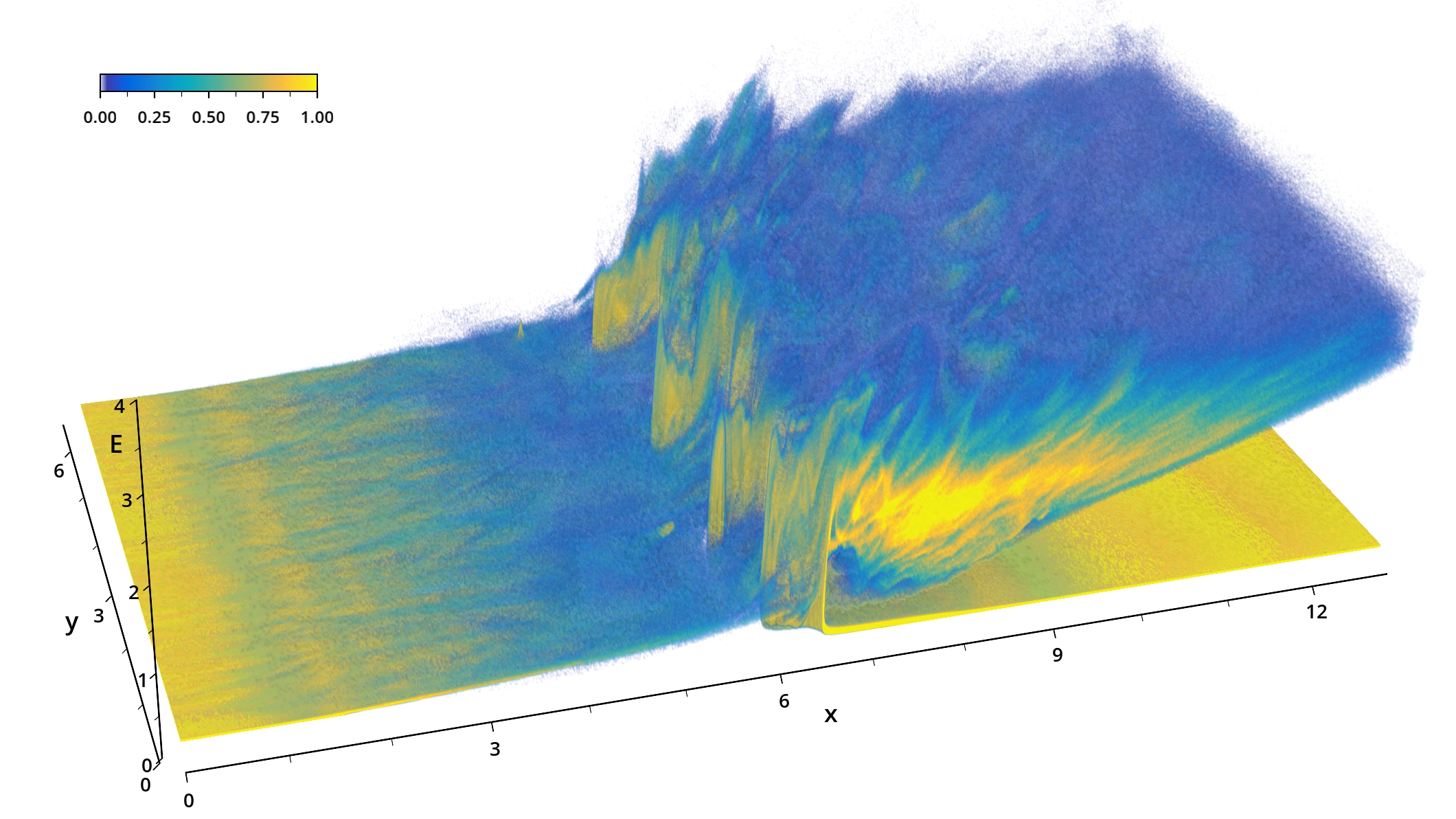}

\includegraphics[width=0.65\textwidth]{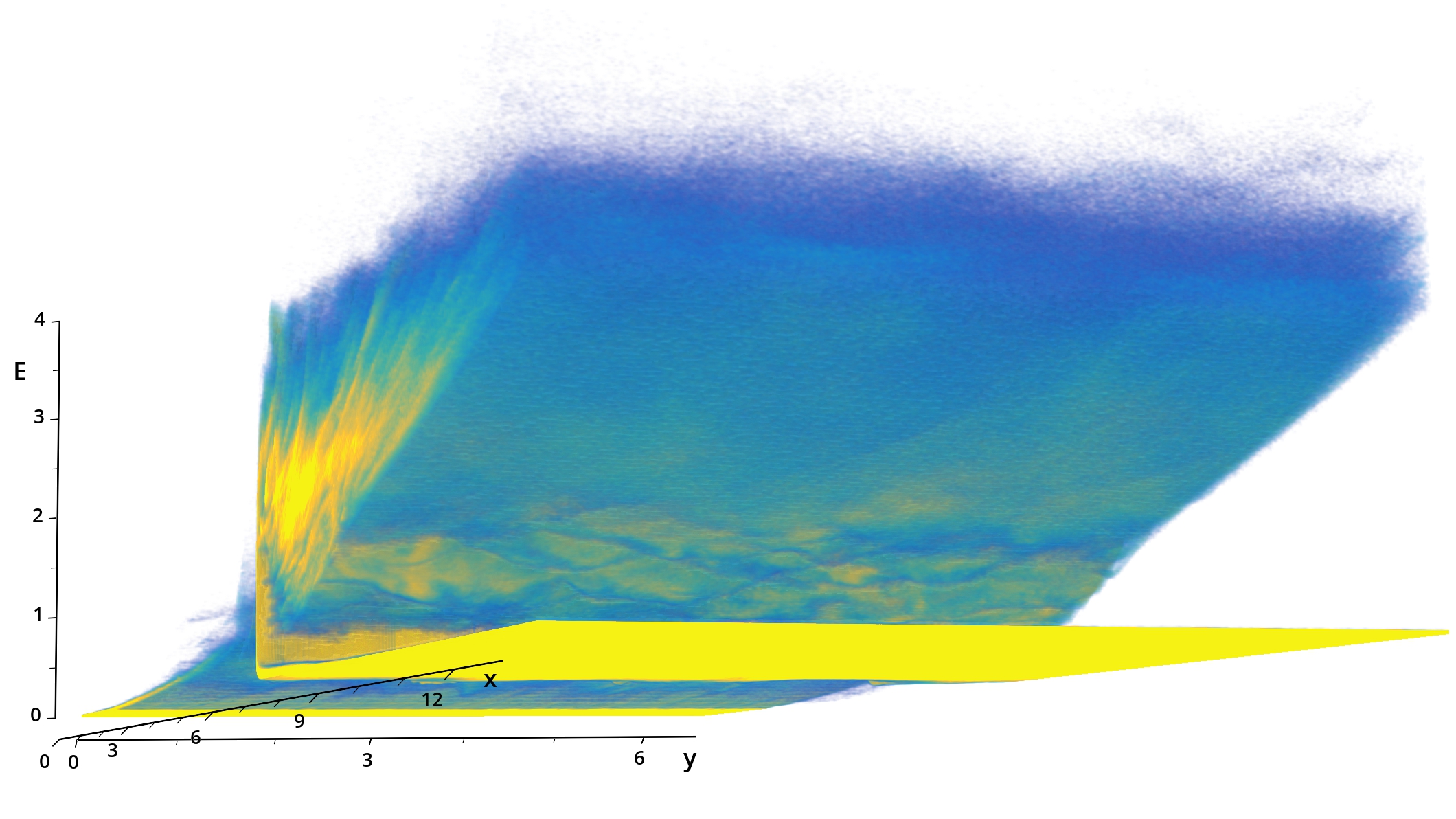}
\caption{Proton energy distribution at the time $t_{sim}=190$ taken from different view directions. We show $f_i(x,y,E)^{1/2}$. Energies are expressed in units of MeV. Multimedia view:}
\label{figure11}
\end{figure*}

Figure~\ref{figure11} presents the proton energy distribution $f_i(x,y,E)^{1/2}$ at the time $t_{sim}=190$ in the same normalization as in Fig.~\ref{figure7}. 
A dilute filamentary proton cloud was left behind by the piston in the interval $x<6$. The mean energy of the protons increases with $x$ for $2 \le x \le 6$, which is characteristic for the trailing end of a solitary wave. A sudden increase of the proton density at low energies is observed where the piston is located. Protons are accelerated at this location and reach energies up to 3 MeV. Thick beams of accelerating protons are observed in the proton density fingers; the in-plane electric field of the piston has funneled protons into these fingers and they get accelerated when they hit the endpoints of the fingers. The piston withstands the ram pressure of the fast and dense protons. The reflected protons leave the finger and move upstream. They spread out rapidly because their interaction with the pleated piston has given them a large thermal velocity spread along $y$. A diffuse and hot distribution of reflected protons is located upstream of the piston. 

Reflected protons are faster than $v_{min}~\approx 1.4~\cdot 10^7$~m/s (1 MeV). Protons with the temperature $T_0$ = 2~keV have a thermal speed $v_{p}={(k_BT_0/m_p)}^{1/2}$ that equals $4.4 \times 10^5$ m/s. The reflected protons thus move at least 30 times faster than $v_p$. Their speed also exceeds 15-fold the ion acoustic speed $c_s={(k_BT_0(\gamma_e+\gamma_p)/m_p)}^{1/2}$ ($\gamma_e = 5/3,\gamma_p=3$ : adiabatic indices of the electrons and protons); their collective interaction with the ambient plasma cannot drive an electrostatic shock that would form during a few inverse proton plasma frequencies. This is because the difference of the electric potential upstream and downstream of the shock is set by the density difference and the electron temperature. Both have an upper limit and so does the potential difference. A shock can only form if the potential difference matches the kinetic energy of the upstream protons measured in the shock frame. This typically limits the maximum speed of an electrostatic shock to a few times the ion acoustic speed. The protons will thus move until their rotation in the upstream magnetic field results in the growth of the fast magnetosonic shock that acts as the boundary between the outer cocoon and the ambient medium in Fig.~\ref{figure1}. We would have to extend the simulation time by an order of magnitude to observe such a shock.\cite{Dieckmann2020}

Finally, we want to test if Rayleigh-Taylor-type instabilities can grow fast enough to explain the observed fingers in the proton density. We assume for simplicity that the piston could maintain a separation of positrons and protons at all times and that the spatial oscillation of the piston grew from a sinusoidal seed perturbation. Winske provides an overview of Rayleigh-Taylor instabilities in magnetized collisionless plasma.~\cite{Winske96} He examines the instability assuming that an electron-ion plasma is placed on top of a spatially uniform unidirectional magnetic field that supports it against the gravitational force. At the time $t=0$ the magnetic pressure balances the plasma pressure. Small displacements of the boundary release gravitational energy, which drives the instability. Several limiting cases exist that yield estimates of the initial growth rate of the instability. We select the case that is closest to the one we observe in our simulation.

Since our instability has fully developed during $eB_0t_{sim}/m_p \approx 0.4$ the protons are essentially unmagnetized. The piston's thickness is about a thermal electron gyro-radius in the ambient plasma ($\approx$ 0.2 in our spatial unit) while the wavelength of the unstable modes is of the order unity at late times (See Fig.~\ref{figure9}). Our instability thus falls into the group of unmagnetized Rayleigh-Taylor instabilities. Its growth rate~\cite{Winske96} is $\gamma = {(kg/A)}^{1/2}$ with an Atwood number  $A=(n_1-n_2)/(n_1+n_2)$ ($n_1, n_2$: densities of the heavy and light fluid) that is about 1 in our case. The gravitational acceleration and the wavenumber of the perturbation along the boundary are $g$ and $k$. 

We estimate the growth rate of the instability using physical units taking a number density for the electrons of the ambient plasma that is $400\, \mathrm{cm}^{-3}$. This value is two orders of magnitude larger than that of the solar wind at the Earth radius. Such a density could be representative for a dense wind of a black hole's companion star. The wavelength 1 of the unstable waves gives $k\approx 2\pi / (10000 \,  \mathrm{m})$ or $k\approx 6\cdot 10^{-4}\mathrm{m}^{-1}$. We assume that upstream protons propagate to the end of the piston, are reflected specularly by its moving boundary and return upstream. Protons are thus accelerated over a distance that is about $2\delta_p$ with the piston thickness $\delta_p \approx 3000 \, m$. A comparison of Fig.~\ref{figure8}(n) and Fig.~\ref{figure9}(a) shows that the piston propagates the distance 2 during the time interval 60 giving a speed $v_p = 2c/60 \approx 10^7$ m/s. Protons thus change their speed from 0 to $v_{min}\approx 1.4 \cdot 10^7$ m/s during the time $\delta_t \approx 2\delta_p / v_p \approx 6\times 10^{-4}$ s. We get a crude estimate of the proton acceleration at the piston $g \approx v_{min}/\delta_t = 2 \times 10^{10}$ m/$\mathrm{s}^2$. The growth rate of the instability is thus $\gamma \approx 3000\, \mathrm{s}^{-1}$ in physical units or $\gamma / \omega_{pi} \approx 0.1$ in normalized ones. Such an instability could easily develop in our simulation. 
 
Can we test if the boundary deforms at such a rate in the simulation? The oscillations at late times have a wavelength $\approx 1$ that exceeds the thickness of the boundary as required by Winske's estimate. Figures~\ref{figure8}(e, f) demonstrate that these oscillations are excited when positrons break through the piston and mix with the protons. If we want to compare our simulation results to Winske's work, we have to examine an oscillation that grows while the piston keeps protons and positrons separated. Figure~\ref{figure8}(a, b) show that this is the case at early times when the when the wavelength of the oscillation is comparable to the boundary thickness. 

\begin{figure}[ht]
\includegraphics[width=\columnwidth]{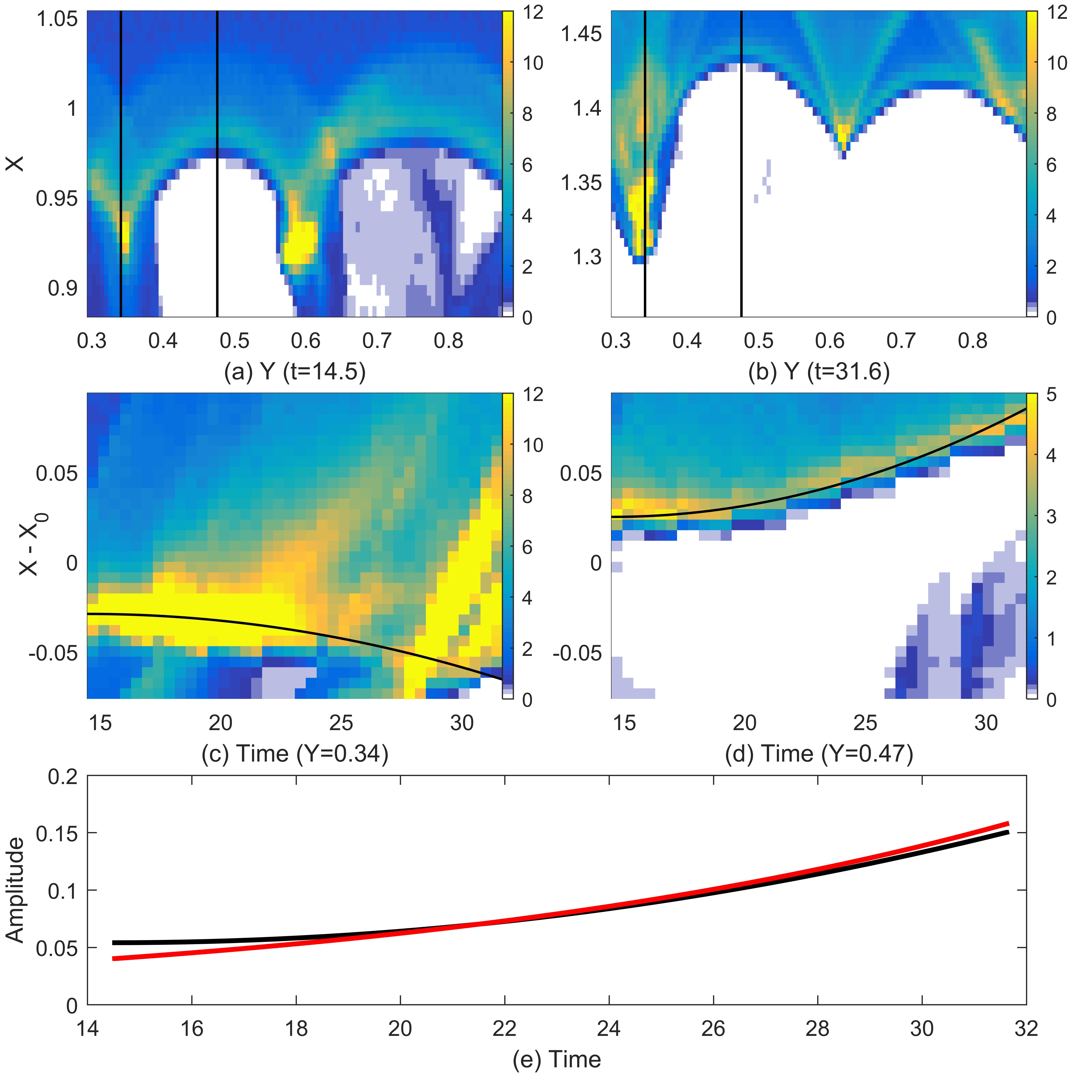}
\caption{Growth rate of the Rayleigh-Taylor-like instability: Panels (a) and (b) show the proton density distributions in a subinterval of the simulation box at the times $t$ = 14.5 and 31.6. Overplotted are the line-outs $y$ = 0.34 and 0.47 where we find extrema of the boundary oscillation. Panels (c) and (d) show the time-evolution of the proton densities along these line-outs. Positions are expressed relative to the reference position $X_0(t) = 0.96 + v_0 (t-14.5)$ where $v_0$ = 0.0234 or $7.4c_s$. The black line in (c) is $s_1(t)=-0.029-{((t-14.5)/90)}^2$ and that in (d) is $s_2(t)=0.025+{((t-14.5)/70)}^2$. Panel (e) compares the difference $s_2-s_1$ (black) of the polynomial fits with the exponential function $s_3(t) = 0.04\exp{(0.08\cdot (t-14.5))}$.}
\label{figure12}
\end{figure}

Figure~\ref{figure12} examines the growth of piston's oscillations at an early time. We display data that has been averaged over 2 cells in each direction. 
The oscillation is no longer sinusoidal at this time as required by a linear stability analysis but we should still be able to get an order-of-magnitude estimate for the growth rate. Figures~\ref{figure12}(a, b) show the proton density distributions at the times $t$ = 14.5 and 31.6. The piston oscillation is periodic at the early time. It is about to double its wavelength at the later time. 
The left oscillation maintains its extrema at the locations $y$ = 0.34 and 0.47 during $14.5 < t < 31.6$. We sample the proton density along these line-outs, we transform them into the moving frame with an origin at $X_0(t) = 0.96 + v_0(t-14.5)$, where $v_0=7.4c_s$ is close to the piston's mean speed. We infer this from the previous observation that the protons, which have been reflected by the piston, reach a peak speed $2v_0$. Figure~\ref{figure12}(c) shows the proton density distribution at $y=0.34$. In time, the high-density region is falling behind the piston as indicated by the polynomial fit $s_1 (t)$. The diagonal high-density structure at times $t>28$ is caused by a proton blob that is catching up with the piston (See Fig.~\ref{figure12}(a) at $y<0.4$ and Fig.~\ref{figure6}(a)). The sharp boundary at $y=0.47$ is fitted well by the polynomial $s_2(t)$. The oscillation amplitude $s_2(t)-s_1(t)$ is plotted in Fig.~\ref{figure12}(e). The amplitude follows closely an exponential curve with the growth rate $\gamma / \omega_{pi}=0.08$. It is close to the one we estimated for piston oscillations with the wavelength 1.   

\section{Discussion}
\label{section4}

We examined the boundary between pair plasma (pair cloud) and cooler uniformly distributed electron-proton (ambient) plasma. The pair cloud was injected at $x=0$ with a uniform density and a mildly relativistic temperature and positive mean speed. Initially, the ambient plasma filled the entire box. A spatially uniform magnetic field permeated the ambient plasma and was oriented orthogonally to the injection direction of the pair cloud and aligned with one of the simulation directions. Our simulation had the purpose to recreate the electromagnetic piston, which was observed in a previous jet simulation,~\cite{Dieckmann2018} with a simplified and computationally cheaper setup. This piston separated the ambient material from the jet material in that simulation and was thus the collisionless counterpart of the contact discontinuity in hydrodynamic jet models. 

We obtained the following results. The expanding pair cloud pushed out the background magnetic field and piled it up ahead of it. We attributed the redistribution of the magnetic field to the diamagnetic current of the hot pair plasma. Its high thermal pressure led to an expansion of the piled-up magnetic field into the ambient plasma trapping its electrons. Their transport with the cloud front induced an electric field. Eventually it became strong enough to accelerate protons. Protons were compressed together with the magnetic field and a solitary fast magnetosonic wave grew. The cloud electrons provided the energy needed to accelerate the protons. Their energy loss resulted in a lower average kinetic energy density of the electrons of the cloud compared to its positrons. The solitary wave broke once the electric field, which was driven by changes in its thermal and magnetic pressure, reached an amplitude that was large enough to reflect the inflowing upstream protons in its rest frame.~\cite{Malkov2016} The saturated solitary wave in our simulation was a discontinuity and not a shock because it separated plasmas with different compositions. We referred to it as a piston. The time it took the piston to form was a few tens of inverse proton plasma frequencies like in previous simulations.\cite{Dieckmann2019,Dieckmann2020}

The pair cloud pushed the piston into the ambient plasma. A boundary, which separates a heavy fluid from a light one that is pushing it, is Rayleigh-Taylor unstable. In the case we considered here, the undular mode with a wavevector parallel to the background magnetic field is destabilized. We estimated the exponential growth rate $\gamma$ of the Rayleigh-Taylor instability for unmagnetized protons based on the work by Winske.~\cite{Winske96} He considered the case where an electron-ion plasma presses against a magnetic field. We found that this instability could grow fast enough to be resolved by our simulation. 

Our simulation confirmed that a Rayleigh-Taylor-like instability grew and deformed the piston. At early simulation times, we found small oscillations of the piston that grew at a rate that was close to that estimated by Winske. Initially, these small oscillations maintained their wavelength and kept protons and positrons separated. 
In time, the nature of the instability changed. Positron fingers could overcome the piston's magnetic field at those locations where positrons had expanded farthest and where the proton density was lowest. Positrons expanded into the electron-proton plasma pushing the weakened piston into the heavy fluid. Broad patches of electromagnetic fields grew around the weakened piston. In time, these fields separated again positrons and protons, which were unique markers for the light and heavy fluids, and the piston reformed. Eventually, fingers formed in the proton density distribution, which were separated by large intervals populated by positrons. 

Magnetic tension stabilizes the undular mode of the Rayleigh-Taylor instability in a magneto-hydrodynamic magnetized plasma,~\cite{Liu2019} which may explain why the growth of the proton fingers stalled or was slow in our simulation. Running the simulation for a much longer time to test if the proton fingers continue to grow is not useful. The plasma conditions at the piston will change once the protons, which have been reflected by the piston, return after their rotation in the upstream magnetic field.\cite{Dieckmann2020} Our simulation has demonstrated for two values of the thermal pressure of the cloud that a piston forms between the pair cloud and the ambient plasma and that it survives as long as the undular mode is involved. Future work has to test the stability of the piston against the interchange mode, where the magnetic field points out of the simulation plane. 

Our simulation demonstrated that stable pistons formed for two different thermal pressures of the pair cloud. A previous 1D study~\cite{Dieckmann2020} showed that the piston can adapt to changes in the ram pressure the upstream ions excert on it. This suggests that such pistons can grow and survive for a wide range of plasma conditions. Their rapid formation time, which gets shorter with increasing plasma densities, also means that they can form much faster than hydrodynamic discontinuities in the collisionless plasma of accretion disk coronae or hot stellar winds. Their robustness and rapid growth makes it likely that they exist in the ultraenergetic plasmas found close to accreting black holes. Flares driven by accretion disk instabilities generate large clouds of electrons and positrons in the disk corona. The piston could keep them separate from the coronal ions with some important consequences. The expansion of the pair cloud across the magnetic field would be slowed down while the pair cloud can still expand along the magnetic field.~\cite{Dieckmann2019} If the pair cloud is permeated by open magnetic field lines, it could expand along them in the form of a jet. The piston would act like the contact discontinuity in hydrodynamic models. However, unlike a discontinuity that is stabilized by particle collisions, the piston is sustained by a strong coherent magnetic field. Its contact with the relativistically hot pair plasma of the inner cocoon would result in electromagnetic wave emissions both during flares in the corona and while the relativistic jet is expanding into the stellar wind of the companion star. 

An intriguing yet speculative aspect of our results is that the conditions that lead to pistons and shocks differ. Radio-synchrotron emissions of relativistic astrophysical jets have been attributed to internal shocks. The magnetic fields that cause such emissions are driven by the beam-Weibel instability and require the collision of pair clouds at mildly relativistic speeds.~\cite{Medvedev99} Shock-generated magnetic fields are usually strong only over a small spatial region of its downstream region.~\cite{Waxman06} They may not be able to produce electromagnetic emissions of relativistic jets at the observed intensities. Our piston forms whenever a pair plasma with a high thermal pressure interacts with a magnetized electron-ion plasma. If an astrophysical jet is composed of pair plasma that flows around slow-moving pockets of ions then pistons would form at the boundaries that separate both. Slow-moving ions could originate from interstellar medium that was ionized by the jet or from ions that entered the jet through its head; the latter is not impermeable for upstream ions if the plasma is collisionless.

\section*{Acknowledgements} MED acknowledges support by a visiting fellowship from the Ecole Nationale Sup\'erieure de Lyon, Universit\'e de Lyon. DF and RW acknowledge support from the French National Program of High Energy (PNHE). The simulation was performed with the EPOCH code financed by the grant EP/P02212X/1 on resources provided by the French supercomputing facilities GENCI through the grant A0070406960 and on resources provided by the Swedish National Infrastructure for Computing (SNIC) at the HPC2N (Ume\aa). 

\section*{Data availability}

Derived data supporting the findings of this study are available from the corresponding author upon reasonable request.

\end{document}